\documentclass[twocolumn,showpacs,preprintnumbers,amsmath,amssymb]{revtex4}

\usepackage{graphicx}
\usepackage{dcolumn}
\usepackage{bm}

\begin{document}


\title{Coexistence of Superconductivity and Antiferromagnetism in Multilayered High-$T_c$ Superconductor HgBa$_2$Ca$_4$Cu$_5$O$_y$: A Cu-NMR Study}

\author{H.~Kotegawa$^{{\rm 1,4},*,\dagger}$, Y.~Tokunaga$^{{\rm 1},\ddagger}$, Y.~Araki$^{\rm 1}$, G.~-q.~Zheng$^{\rm 1}$, Y.~Kitaoka$^{\rm 1,4}$ \\
K.~Tokiwa$^{\rm 2,4}$, K.~Ito$^{\rm 2}$, T.~Watanabe$^{\rm 2,4}$, A.~Iyo$^{\rm 3,4}$, Y.~Tanaka$^{\rm 3,4}$, and H.~Ihara$^{\rm 2,3,4}$}
 \altaffiliation{$\dagger$ Present Address: Department of Physics, Okayama University. Tsushima-naka 3-1-1, Okayama 700-8530 Japan} 
 \email{kotegawa@science.okayama-u.ac.jp}
 \altaffiliation{$\ddagger$ Present Address : Japan Atomic Energy Research Institute, Tokai-mura, Ibaraki 319-1195, Japan}
 \email{tokunaga@popsvr.tokai.jaeri.go.jp}
\affiliation{%
$^{\rm 1}$Department of Physical Science, Graduate School of Engineering Science, Osaka University, Toyonaka, Osaka 560-8531, Japan \\
$^{\rm 2}$Department of Applied Electronics, Science University of Tokyo, Yamazaki, Noda, Chiba 278-8510, Japan \\
$^{\rm 3}$National Institute of Advanced Industrial Science and Technology (AIST), Umezono, Tsukuba 305-8568, Japan \\
$^{\rm 4}$Core Research for Evolutional Science and Technology (CREST) of the Japan Science and Technology Corporation (JST), Kawaguchi, Saitama 332-0012, Japan \\
}%

\date{\today}

\begin{abstract}
We report a coexistence of superconductivity and antiferromagnetism in five-layered compound HgBa$_2$Ca$_4$Cu$_5$O$_y$ (Hg-1245) with $T_c=108$ K, which is composed of two types of CuO$_2$ planes in a unit cell; three inner planes (IP's) and two outer planes (OP's). The Cu-NMR study has revealed that the optimallydoped OP undergoes a superconducting (SC) transition at $T_c=108$ K, whereas the three underdoped IP's do an antiferromagnetic (AF) transition below $T_N\sim$ 60 K with the Cu moments of $\sim (0.3-0.4)\mu_B$. 
Thus bulk superconductivity with a high value of $T_c=108$ K and a static AF ordering at $T_N=60$ K are realized in the alternating AF and SC layers.
The AF-spin polarization at the IP is found to induce the Cu moments of $\sim0.02\mu_B$ at the SC OP, which is the AF proximity effect into the SC OP.

\end{abstract}

\maketitle

\section{INTRODUCTION}

There remain a number of underlying issues to be resolved in high-$T_c$ cuprates.
One of underlying issues is an interplay of antiferromagnetism and superconductivity in the antiferromagnetic (AF) - superconducting (SC) phase boundary, near vortex cores under magnetic field, and AF-SC alternate layered structures.
Lake {\it et al.} reported that the AF correlations  are induced in vortex cores and extend over the cores into the SC region in La$_{2-x}$Sr$_{x}$CuO$_{4}$ under the magnetic field, that is, an AF proximity effect into SC state. \cite{Lake}
This result is supported by some theoretical approaches.\cite{Hu,Demler}
In these theoretical predictions based on the SO(5) symmetry model, \cite{Zhang} each AF and SC fluctuation can extend into other region, when both the states come across.
Recently, however, Bozovic {\it et al.} showed that the superconductivity does not mix into the AF insulator in the superconductor-insulator-superconductor heterostructures realized by stacking each layer of SC La$_{1.85}$Sr$_{0.15}$CuO$_{4}$ and AF La$_{2}$CuO$_{4}$. \cite{Bozovic} Thus, these issues have not been settled yet.

Multilayered high-$T_c$ cuprates, which have more than three CuO$_2$ planes in a unit cell, exhibit very unique magnetic and SC properties 
because they include two types of CuO$_2$ planes. As indicated in Fig.~1, 
an outer CuO$_2$ plane (OP) has a pyramidal five-oxygen coordination, whereas an inner plane (IP) has a square four-oxygen one.
Note that the IP$^*$ is the middle plane of the three IP's as shown in the figure.
Nuclear-magnetic-resonance (NMR) experiments revealed that 
the OP and the IP differ in the doping level.\cite{Trokiner,Statt,Piskunov,Tokunaga0}
We reported unusual magnetic and SC characteristics in multilayered CuO$_2$ planes in Hg and Cu-based high-$T_c$ cuprates through $^{63}$Cu-NMR measurements. \cite{Kotegawa} The Knight shift ($^{63}K$) at the OP and the IP exhibits different characteristic temperature ($T$) dependence, consistent with its own doping level.   It was shown that the doping level $N_h$(OP) at the OP is larger than $N_h$(IP) at the IP for all the systems and its difference $\Delta N_h=N_h$(OP)$-N_h$(IP) increases as either a total carrier content or $n$ increases. At $\Delta N_h$'s exceeding a critical value,  the respective SC transitions do not simultaneously set in  at the IP and the OP. \cite{Tokunaga}
Some theoretical approaches predict the effect induced by the carrier inhomogeneity. \cite{Mori,Chakravarty}
\begin{figure}[htbp]
\includegraphics[width=.7\linewidth]{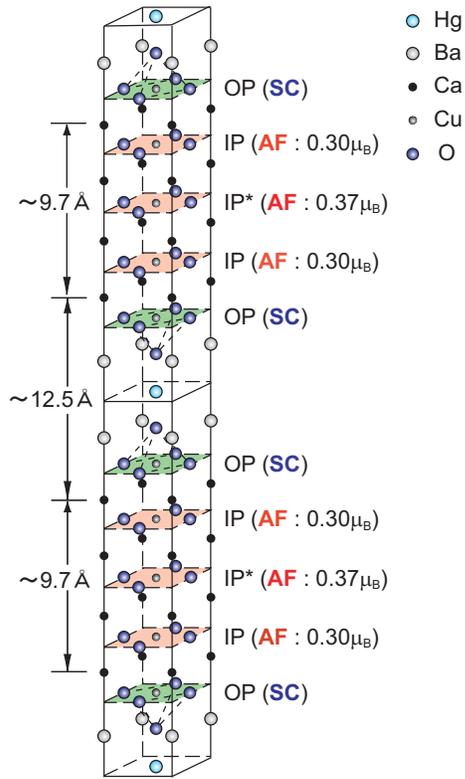}
\caption[]{\protect The crystal structure of Hg-1245 ($a=3.850~\AA$, $c=22.126~\AA$) (Ref.6). The OP undergoes the SC transition at $T_c=108$ K, whereas the three underdoped IP's do an AF transition below $T_N\sim$ 60 K with the respective Cu moments of $\sim 0.30\mu_B$ and 0.37$\mu_B$ at the IP and the IP$^*$.
}
\end{figure}

In this paper, we report $^{63,65}$Cu-NMR study on Hg-1245 which evidences a coexistence of AF order at the IP's and IP$^*$ and bulk superconductivity at the OP. Note that the IP$^*$ is the middle plane of the three IP's as shown in Fig.~1.
Measurements of the Knight shift $^{63}K$, the nuclear-spin-lattice-relaxation rate ($1/T_1$) and the internal field ($H_{\rm int}$) of $^{63,65}$Cu have revealed that the OP undergoes a bulk SC transition below $T_c=108$ K and the IP* and IP order antiferromagnetically below $T_N\sim$ 60 K with the Cu moments of 0.37$\mu_B$ and 0.30 $\mu_B$, respectively.

\section{EXPERIMENTAL DATA AND DISCUSSION}

Polycrystalline sample was prepared by the high-pressure synthesis technique as described elsewhere. \cite{Tokiwa}
Powder x-ray-diffraction experiment indicates that the sample consists of almost a single phase, but includes a small fraction of Hg-1234.\cite{Tokiwa} 
A SC transition temperature of $T_c=108$ K was determined from an onset $T$ below which diamagnetic signal appears in dc susceptibility as shown in Fig.~2(a). 
For NMR measurements, the powdered sample was aligned along the $c$ axis at $H=16$ T.
The NMR experiment was performed by the conventional spin-echo method at 174.2 MHz ($H\sim$ 15.3 T).
\begin{figure}[htbp]
\includegraphics[width=.9\linewidth]{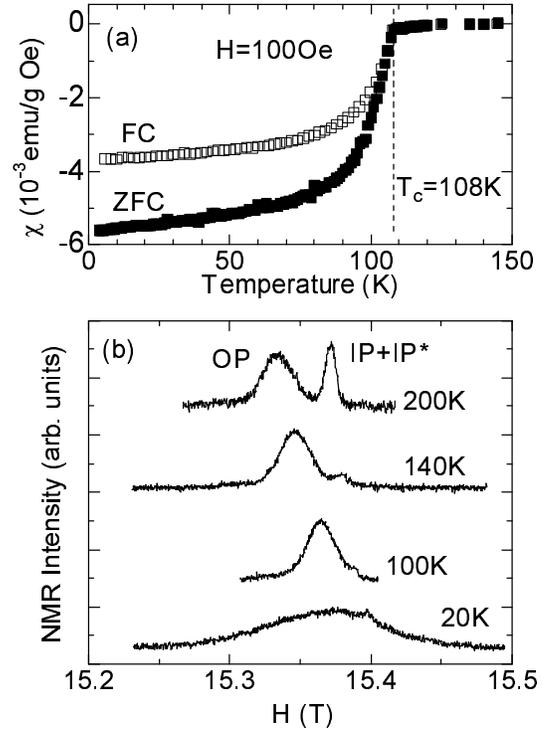}
\caption[]{\protect (a) $T$ dependence of dc susceptibility with field cooling (FC) and zero-field cooling (ZFC). A clear diamagnetic signal can be observed below $T_c=108$ K.
(b) The NMR spectra for $H\perp c$ at $T=200$, 140, 100, and 20 K. The NMR signals at the IP and the IP$^*$ disappear below $\sim 150$ K. The OP's signal becomes significantly broader at temperatures lower than $T_{N}\sim 60$ K.
}
\end{figure}

Figure~2(b) shows NMR spectra for $H \perp c$. At 200 K, two well-separated peaks  arise from the OP and the IP.  The assignment of NMR spectrum to the OP and the IP  was already reported in the previous literatures.\cite{Kotegawa,Zheng2}
The NMR spectrum at the IP exhibits a sharper spectral width with a smaller Knight shift than those at the OP. The spectral width at the IP is estimated to be $\sim 50$ Oe for $H\parallel c$, comparable to the $\sim 60$ Oe for YBa$_2$Cu$_3$O$_7$ under $H\sim 15$ T, which is the narrowest among high-$T_c$ cuprates to date.  This ensures that the IP is rather homogeneously doped. The spectra at the IP and the IP$^*$ overlap each other, suggesting that their local doping levels  are  not so much different. 
The  NMR signals at the IP and the IP$^*$ disappear due to their short relaxation time below $T\sim150$ K.

Figure~3 indicates the $T$ dependence of $^{63}K_{ab}$ at the IP's and the OP for $H\perp c$ .  In general, $K(T)$ consists of the $T$-independent orbital part, $K_{orb}$, and the $T$-dependent spin part, $K_s(T)$, that is proportional to the uniform susceptibility $\chi_s$. $K_{s}$(OP) decreases below $T^*\sim 160$ K, followed by a rapid decrease around $T_c=108$ K which is indicative of the bulk superconductivity at the OP, as also confirmed  in the measurement of $1/T_1T$.
The behavior of $K_{ab}(T)$ suggests that the OP is almost optimally doped from a comparison with the previous study. \cite{Kotegawa}
In the inset of Fig.~3, $K_{ab}(T)$ vs $K_c(T)$ plots are presented at the OP and the IP's. The spin part in the measured shift $K_{s,\alpha}(T)$ at the CuO$_2$ plane is expressed as following, \cite{Mila}
\[
K_{s,\alpha}=(A_{\alpha}+4B)\chi_{s,\alpha}\ \ \ (\alpha=a, b, {\rm and}~c ~{\rm axis}),
\]
where $A_\alpha$ and $B$ are the on-site and the supertransferred hyperfine-coupling constants. $A_\alpha$ is anisotropic, mainly originating from the dipole and the spin-orbit interactions for Cu-$3d$ orbitals, and $B$ is isotropic, originating from the Cu($3d_{x^2-y^2}$)-O($2p$)-Cu($4s$) covalent bonding. $\chi_{s,\alpha}$ is the spin susceptibility.
From a linear relation in the figure, $(A_c+4B)/(A_{ab}+4B)\sim$ 0.267 and  
0.379 are estimated at the IP's and the OP, respectively.
By assuming $A_c \sim -170$ kOe/$\mu_B$ and $A_{ab}\sim 37$ kOe/$\mu_B$ in YBa$_2$Cu$_3$O$_7$, \cite{Millis,Monien,Imai}  the respective values of $B$ at the IP's and the OP in Hg-1245 are estimated as $B({\rm IP})\sim 61$ kOe/$\mu_B$ and $B({\rm OP})\sim 74$ kOe/$\mu_B$. These values of $B$ are larger than the typical value of $\sim 40$ kOe/$\mu_B$ obtained in La$_{2-x}$Sr$_x$CuO$_4$, YBa$_2$Cu$_3$O$_7$, and YBa$_2$Cu$_4$O$_8$, \cite{Millis,Monien,Imai,Zheng,Monien2,Ishida} suggesting that the Cu($3d_{x^2-y^2}$)-O($2p$)-Cu($4s$) covalent bonding in Hg-1245 is stronger than in the La- or Y-based systems.
\begin{figure}[htbp]
\includegraphics[width=.9\linewidth]{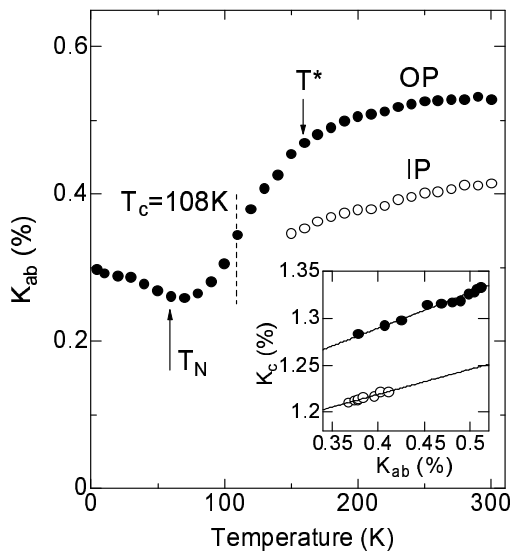}
\caption[]{\protect $T$ dependence of Knight shift $K_{ab}$ for $H\perp c$. The inset of Fig.~3(b) shows $K_{ab}$ vs $K_c$ plots at the IP's and the OP with the temperature as an implicit parameter. These plots allow us to estimate the supertransferred hyperfine-coupling constants $B({\rm IP})\sim61$ kOe/$\mu_B$ and $B({\rm OP})\sim74$ kOe/$\mu_B$, respectively.
}
\end{figure}

Next, we present firm evidence for the occurrence of AF ordering at the IP and the IP$^*$. 
Figure~4 shows $^{63,65}$Cu-NMR spectra at $H=0$ and $T=1.4$ K.
Four and two peaks are observed in the frequency ranges of $f=55-110$ and $10-40$ MHz, respectively.
The nuclear quadrupole frequencies at the IP and the OP's, $\nu_{Q}$(IP) and $\nu_{Q}$(OP), are estimated from the NMR experiments at high $T$ as $^{63}\nu_{Q}({\rm IP})=8.37$ MHz and $^{63}\nu_{Q}({\rm OP})=16.05$ MHz, respectively (not shown).
Therefore, all these spectra are affected by the presence of internal field $H_{\rm int}$ associated with the onset of AF order. 
\begin{figure}[htbp]
\includegraphics[width=\linewidth]{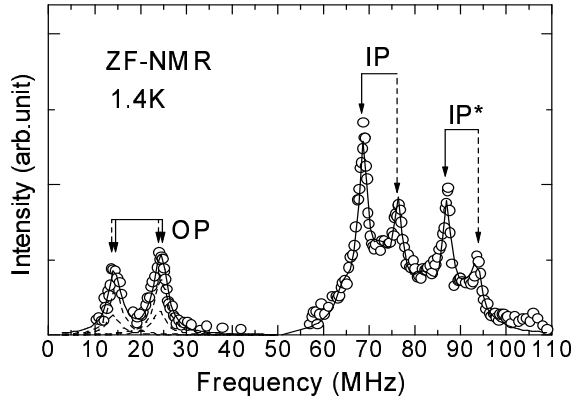}
\caption[]{\protect  A zero-field $^{63,65}$Cu-NMR spectra at 1.4 K. Four peaks in $f=55-110$ MHz, which consist of two sites (IP and IP$^*$) and two isotopes [$^{63}$Cu (solid arrow) and $^{65}$Cu (dashed arrow)], are observed. The spectra in $f=10-40$ MHz correspond to the OP. The solid lines are the simulation calculated by using $^{63}\nu_{Q}({\rm IP})=8.37$ MHz, $^{63}\nu_{Q}({\rm OP})=16.05$ MHz, and $H_{\rm int}$ along $ab$ plane. Each Cu moment is estimated as $M({\rm IP})\sim0.30\mu_B$ and $M({\rm IP^*})\sim0.37\mu_B$ (see text). The internal field of $H_{\rm int}\sim 0.54$ T exists even at the SC OP.
}
\end{figure}

The nuclear Hamiltonian ${\cal H}={\cal H}_{\rm Q}+{\cal H}_{\rm Z}$ at $H=0$ below $T_N$ is described in terms of the Zeeman interaction due to $H_{\rm int}$ and the nuclear electric quadrupole interaction as follows:
\[
{\cal H}_{\rm Z} = -\gamma_n\hbar \left\{ \frac{1}{2}H_{\perp}(I_{+}+I_{-})+H_{\parallel} I_z \right\},
\]
where $H_{\perp}$ and $H_{\parallel}$ are the respective components perpendicular and parallel to the $c$-axis and $\gamma_n$ is the Cu nuclear gyromagnetic ratio, and
\[
{\cal H}_Q=\frac{e^2qQ}{4I(2I-1)}\{[3I_z^2-I(I+1)]+\eta(I_x^2-I_y^2)\},
\]
where $\eta$ is the asymmetry parameter of electric-field gradient.
Here, note that the quadrupole frequency $h\nu_{\rm Q} \equiv 3e^2qQ/2I(2I-1)$ and $\eta \sim 0$.

The spectra observed in $f=55-110$ MHz correspond to the case for $\nu_{\rm Q}\ll H_{\rm int}$ due to the AF order below $T_{N}\sim$ 60 K.
Four peaks are understood as the central peaks ($1/2\leftrightarrow -1/2$ transition) of $^{63,65}$Cu at the IP and the IP$^*$.  A ratio of $^{63,65}$Cu-NMR intensity at low frequency to high frequency ($I_L/I_H\sim 2$) suggests that the two peaks at low (high) frequencies  arise from the IP (IP$^*$).  The satellite peaks ($\pm 1/2 \leftrightarrow \pm 3/2$ transition) due to the electric quadrupole interaction are not well resolved. By incorporating this intensity ratio $I_L/I_H\sim 2$, $^{63}\nu_{Q}({\rm IP})=8.37$ MHz, $^{63}Q/^{65}Q\sim1.08$, and $^{63}\gamma_n/^{65}\gamma_n\sim0.93$, the NMR spectra at the IP and the IP$^*$are simulated as the solid line in the figure, giving rise to the respective values of $H_{\rm int}({\rm IP})=6.1$ T and $H_{\rm int}({\rm IP^*})=7.7$ T.  These values at the IP and the IP$^*$ allow us to estimate the Cu moments  $M({\rm IP})\sim 0.30\mu_B$ and $M({\rm IP^*})\sim 0.37\mu_B$ by using a hyperfine-coupling constant $(A_{ab}-4B)\sim -207$ kOe/$\mu_B$ where $B({\rm IP})\sim 61$ kOe/$\mu_B$. These Cu moments are one-half smaller than $0.64\mu_B$ estimated in  La$_2$CuO$_4$.\cite{Tsuda}  
We remark that the $N_h({\rm IP})$ and $N_h({\rm IP^*})$ are tentatively estimated as $[5\delta_{av}-2N_h({\rm OP})]/3\sim 0.057\pm0.02$ by using an average hole content $\delta_{av}=0.12$ evaluated from a Hall measurement.\cite{Tokiwa3} Note that $N_h({\rm OP})=0.212-0.217$ was estimated via the systematic experimental relation between the $N_h$ and the $K_s$ at room temperature argued in the literature, \cite{Kotegawa} and hence it is shown that the OP is optimally doped. Here we assumed $K_{orb}$ for Hg-1245 to be $0.19-0.20$ from a comparison with other multilayered cuprates.\cite{Zheng2,Julien}

The spectra in $f=10-40$ MHz suggest the case for $\nu_{\rm Q}\sim 16$ MHz $\simeq H_{\rm int}$. Actually, the calculated spectra to be consistent with the experiment are indicated in the figure, allowing us to estimate $\nu_{\rm Q}\sim 16$ MHz and $H_{\rm int}=H_{\perp}\sim 0.54$ T. These spectra are hence assigned as arising from the OP. 
This $H_{\rm int}$ of 0.54 T is far larger than the calculated dipole field in the OP of $\sim 70$ Oe, which is induced by the Cu moments of $M({\rm IP})\sim 0.30\mu_B$ and $M({\rm IP^*})\sim 0.37\mu_B$.  $H_{\rm int}{\rm (OP)}\sim 0.54$ T corresponds to the Cu moments of $\sim0.02 \mu_B$.

\begin{figure}[htbp]
\includegraphics[width=0.95\linewidth]{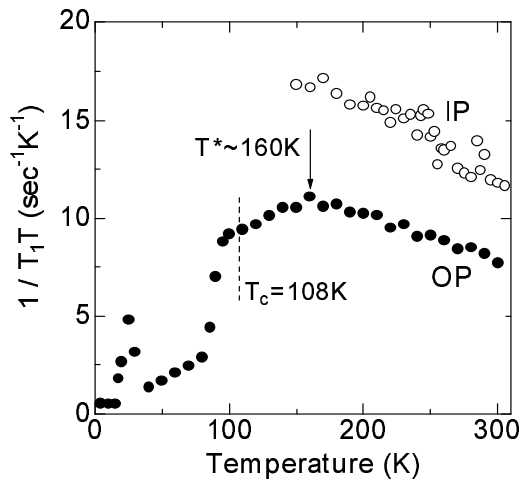}
\caption[]{\protect $T$ dependence of $1/T_1T$ for $H\parallel c$. $T_1$(OP) below $\sim90$ K is the long component in the recovery curve. The pseudogap behavior is observed at the optimallydoped OP below $T^* \sim 160$ K. On the other hand, the underdoped IP does not show any pseudogap indication, revealing that the low-energy spectral weight in $\chi({\bf q}={\bf Q}, \omega)$ is critically enhanced around $\omega\sim 0$ toward AF ordering at $T_N \sim 60$ K.
}
\end{figure}

Figure~5 indicates the $T$ dependence of $1/T_1T$ of $^{63}$Cu at the IP's and the OP for $H\parallel c$. Remarkably, $1/T_1T$(OP) exhibits a pseudogap behavior below $T^*\sim$ 160 K. $1/T_1$(OP) is distributed below $\sim 90$ K. In the normal state, a recovery curve of nuclear magnetization is consistent with a theoretical one for determining a single value of  $T_1$ as seen in Fig.~6(a).\cite{Narath} 
Below $\sim 90$ K, however, a short component in the recovery curve is observed as presented in Fig.~6(b).  A tentative fitting to the curve, which is indicated by a solid line, allows us to estimate a short and a long component in $T_1$.
Their $T$ dependencies are shown in Fig.~7, where the short components are presented by open square.

\begin{figure}[htbp]
\includegraphics[width=.8\linewidth]{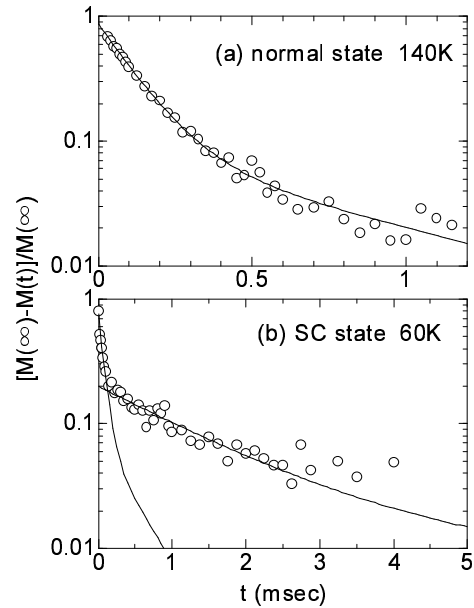}
\caption[]{\protect The recovery curve of nuclear magnetization for determining $T_1$ at the OP at (a) the normal and (b) the SC state, respectively.  A short component in the recovery curve is observed as presented below $\sim90$ K in the SC state.
}
\end{figure}

Generally in the SC mixed state under magnetic field, the short component in $T_1$ is believed  to arise due to the presence of the {\it normal state} in vortex cores.  But the large fraction of the short component reaching $\sim80 \%$ is unusual, giving rise to almost a same fraction as the short component  at $H=0$ indicated by open triangle in the figure.
These results ensure that the short component dose not arise due to the presence of  vortex cores,  but originates from the unexpected relaxation process at the SC OP.
Thus,  some low-lying magnetic excitations  survive  at the SC OP even though the $d$-wave superconductivity is formed  well below $T_c=108$ K.
The short component indicates two peaks at $T_{N}\sim 60$ K and $T_{?}\sim 25$ K, whereas the long component indicates a peak at $T_{?}\sim 25$ K. $T_N\sim 60$ K is indicative of an onset of AF ordering at the IP's, corroborated by the increase of $K_s({\rm OP})$. Recent muon spin resonance measurement also evidences an AF ordering below $\sim 60$ K in this material. \cite{Tokiwa2} $T_{?}\sim 25$ K might be related to the occurrence of $H_{\rm int}$(OP) because both the long and short components show peaks.

\begin{figure}[htbp]
\includegraphics[width=0.95\linewidth]{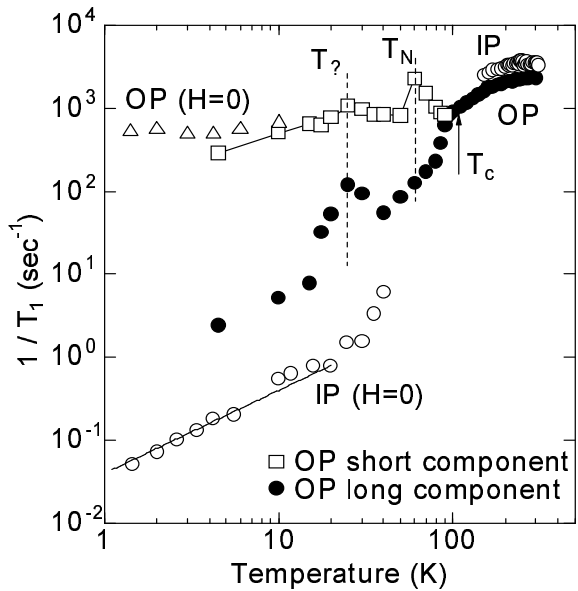}
\caption[]{\protect $T$ dependence of  $1/T_1$ for $H\parallel c$. $1/T_1$ below $T_N$ can be measured at zero field. $1/T_1$(OP) shows the peak at $T_N \sim 60$ K associated with AF ordering at the IP. $1/T_1$(IP) shows a $T_1T \sim {\rm const}$ relation far below $T_N$, indicating that the IP is metallic. $[T_1({\rm OP})/(T_1({\rm IP})]\sim 10^{-3}$ at low $T$ shows the existence of low-lying magnetic excitations inherent to the OP, which is associated with a possible interplay with the superconductivity.
}
\end{figure}

On the other hand, $1/T_1T$(IP) increases monotonically, whereas $K_s$(IP) decreases with decreasing $T$ down to 150 K.  Instead of the pseudogap, unexpectedly, the NMR signals at the IP and the IP$^*$ disappear below $\sim 150$ K. This is because the low-energy spectral weight in dynamical response function $\chi({\bf q=Q}, \omega)$ is critically enhanced around $\omega\sim 0$. Here {\bf Q} is the AF wave vector $(\pi/a,\pi/a)$. 
Eventually, the IP's order antiferromagnetically below $T_{N}$ as evidenced from the zero-field AF NMR experiment that probes the Cu moments of $M({\rm IP})\sim 0.30\mu_B$ and $M({\rm IP^*})\sim 0.37\mu_B$. 
The measurement of $T_1$ at the IP reveals a behavior of $T_1T\sim$ const below $\sim 20$ K  in the SC state at $H=0$ as seen in Fig.~7.
In magnetically ordered metals, a  nuclear-spin-relaxation process is mediated by the  interaction between nuclear spins and conduction electrons via spin-wave excitations, which is called as the Weger mechanism leading to a behavior of $T_1T= {\rm const}$ at low $T$.\cite{Wegar,Matsumura}
Thus the IP is suggested to be not in an insulating regime but a metallic one, which is consistent with the estimated hole content of $\sim 0.057\pm0.02$.
The small value of $T_1T= {\rm const}$ indicates that a large gap opens in the magnetic excitation, indicating that the AF ordering in the IP is in a static regime.

When the SC state is closely faced to the AF state realized in the doped CuO$_2$ plane, it is not obvious to what extent the superconductivity is affected because of the presence of AF state.  Hg-1245 is a good candidate to address this issue.  Both the measurements of $T_1$ and Knight shift evidence that the OP is in the SC state.
If the AF-spin polarization at the IP induced the $H_{\rm int}{\rm (OP)}=0.54$ T via the hybridization between $4s$(IP) and $4s$(OP) and/or $3d_{3z^2-r^2}$(OP) [not $3d_{x^2-y^2}$(OP)], a ratio of $[T_1({\rm OP})/T_1({\rm IP})] \sim [H_{\rm int}({\rm IP})/H_{\rm int}({\rm OP})]^2\sim  [6.1T/0.54T]^2\sim 10^2$ would be expected.  It is, however, surprising that the $T_1$ at the SC OP is $10^3$ times shorter than at the antiferromagnetically ordered IP at low $T$ far below $T_N$, being $[T_1({\rm OP})/(T_1({\rm IP})]\sim 10^{-3}$. This result demonstrates the existence of low-lying magnetic excitations inherent to the OP associated with a possible interplay with the superconductivity.  It suggests that the weak AF order with small moment 0.02 $\mu_B$ is responsible for the low-lying magnetic fluctuations at the OP and coexists with the SC state at the OP.  We note that this coexistence seems to be similar to the phenomenon near vortex cores where the AF correlations originating from the vortex cores extend over the cores into the SC region.\cite{Lake}  
On the other hand, it is quite interesting issue whether the SC order parameter exists at the metallic AF IP in Hg-1245, but this is a future issue, because the NMR signal from the  
AF IP disappears in $T=40-150$ K.

\begin{figure}[htbp]
\includegraphics[width=.8\linewidth]{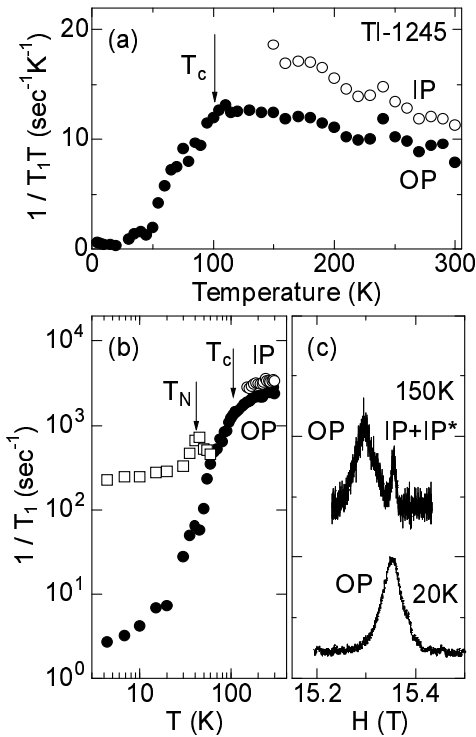}
\caption[]{\protect $T$ dependence of (a) $1/T_1T$ and (b) $1/T_1$ for $H\parallel c$ in Tl-1245, which is slightly overdoped compared with Hg-1245. The short component of $T_1$(OP) shows the peak at $\sim 45$ K, indicative of AF ordering at the IP. $T_?$ is absent in Tl-1245. Figure (c) presents NMR spectra at 150 K and 20 K. No change is observed in the linewidth of the OP signal, indicating that the internal field at the OP is quite tiny, in contrast with Hg-1245.
}
\end{figure}

Finally we mentioned some results of TlBa$_2$Ca$_4$Cu$_5$O$_y$ (Tl-1245) with $T_c=100$ K, which is slightly much overdoped than Hg-1245.
As shown in Fig.~8(a), the $1/T_1T$(OP) of Tl-1245 does not show the pseudogap behavior, indicating that the OP is in overdoped regime. $1/T_1T$(IP) increases monotonically with decreasing $T$ and the signals of the IP and the IP* disappear below $\sim140$ K as well as Hg-1245. The low-lying magnetic excitations are induced at the SC OP also in Tl-1245, and thus $T_1$(OP) distributes below $\sim 60$ K. Its short components show a peak at $\sim45$ K, corresponding to the peak of $T_N\sim60$ K of Hg-1245, as shown in Fig.~8(b). This $T_N \sim 45$ K at the IP of Tl-1245 suggests that the IP of Tl-1245 has somewhat much carrier content than that of Hg-1245. Interestingly, however, the second anomaly corresponding to $T_?$ disappears in Tl-1245, and the line width of the OP does not change between 150 K and 20 K, which is quite contrast to Hg-1245 as shown in Figs.~8(c) and 2(b).  This ensures that the internal field  at the OP of Tl-1245 is quite tiny even below $T_N \sim 45$ K at the IP, which is different from $H_{\rm int}{\rm (OP)} \sim 0.54$ T of Hg-1245.  These things imply that $H_{\rm int}{\rm (OP)} \sim 0.54$ T is induced below $T_? \sim 25$ K in Hg-1245.

\section{SUMMARY}

In summary, the $^{63,65}$Cu-NMR measurements have unraveled that 
the disparate electron phases emerge at the outer two CuO$_2$ planes and the inner three ones in HgBa$_2$Ca$_4$Cu$_5$O$_y$.  The $T$ dependencies of Knight shift and $1/T_1$ have revealed that the optimallydoped OP undergoes the bulk SC transition at $T_c=108$ K and the underdoped IP's do the AF transition below $T_{N}\sim 60$ K without any indication of pseudogap. The zero-field $^{63,65}$Cu-NMR experiments at low $T$ have provided firm evidence that the respective AF moments at the IP and the IP$^*$ are $M({\rm IP})\sim0.30\mu_B$ and $M({\rm IP^*})\sim0.37\mu_B$. The bulk superconductivity with the high value of $T_c=108$ K and the static AF ordering at $T_N=60$ K take place even though the AF and SC layers are alternatively stacked with the respective thickness being comparable with $\sim 9.7~\AA$ and $\sim 12.5~\AA$. The AF-spin polarization at the IP is found to induce the Cu moments of $\sim 0.02\mu_B$ at the OP that fluctuates faster than the AF moment at the IP does, evidencing the AF proximity effect into the SC OP.

\begin{acknowledgments}
The authors are grateful to Kenji Ishida for his helpful discussions and Shotaro Morimoto for experimental supports in dc susceptibility measurement.
This work was supported by the COE Research (Grant No.10CE2004) in Japan NEXT Grant. 
One of authors (H.K.) has been supported by {\it JSPS Research Fellowships for Young Scientists}.
\end{acknowledgments}

\end{document}